\documentclass{INTERSPEECH2023}

\interspeechcameraready

\title{HOW TO (VIRTUALLY) TRAIN YOUR SPEAKER LOCALIZER}
\name{Prerak Srivastava$^{1}$, Antoine Deleforge$^{1}$, Archontis Politis$^{2}$, Emmanuel Vincent$^{1}$}

\address{$^{1}$Universit\'e de Lorraine, CNRS, Inria, Loria, F-54000 Nancy, France\\$^{2}$Audio and Speech Processing Research Group, Tampere University, Finland
}
\email{$^{1}$\{prerak.srivastava, antoine.deleforge, emmanuel.vincent\}@inria.fr \\
$^{2}$archontis.politis@tuni.fi}

\begin{document}

\maketitle
 
\begin{abstract}
% 1000 characters. ASCII characters only. No citations.deep learning
Learning-based methods have become ubiquitous in speaker localization. Existing systems rely on simulated training sets for the lack of sufficiently large, diverse and annotated real datasets. Most room acoustics simulators used for this purpose rely on the image source method (ISM) because of its computational efficiency.  
This paper argues that carefully extending the ISM to incorporate more realistic surface, source and microphone responses into training sets can significantly boost the real-world performance of speaker localization systems. It is shown that increasing the training-set realism of a state-of-the-art direction-of-arrival estimator yields consistent improvements across three different real test sets featuring human speakers in a variety of rooms and various microphone arrays. An ablation study further reveals that every added layer of realism contributes positively to these improvements. 
\end{abstract}

\section{Introduction}
\label{sec:intro}

%Deep-learning based methods for processing audio recordings such as speech enhancement \cite{wang2020complex}, source separation \cite{cord2022monaural}, acoustic analysis \cite{srivastava2022realistic}, and sound source localization and tracking \cite{grumiaux2022survey} require large amounts of training data. Due to the difficulty of obtaining enough real data with enough diversity, these training data are typically simulated, an approach called \textit{virtually supervised learning} in, e.g., \cite{foy2021mean,srivastava2022realistic}. For those applications on which the effect of reverberation is detrimental, such as localization or multi-microphone speech enhancement, the use of room acoustics simulators is widespread since it allows physical modeling of the room impulse response (RIR) from the source to the receivers, which includes the inter-channel differences utilized by the methods as well as reverberation. Depending on the type of simulation method used, diverse data can be generated in terms of source and receiver position, acoustic material characteristics, or room geometry, which is necessary for the generalization of the methods to a wide range of real acoustic scenes.

Far-field deep learning based speech processing systems often require large amount of training data, as demonstrated by their application to various tasks such as speech recognition \cite{gusev2020deep,huang2019intel}, speaker localization \cite{xue2020sound,grumiaux2022survey}, speech enhancement \cite{chen2022multi,rao2021conferencingspeech} and diarization \cite{horiguchi2020end,ryant2020third}. For increased generalization, the study in \cite{vincent2017analysis} suggest the use of extensive training sets that cover the range of variability found in real test sets. Due to the difficulty of obtaining enough real data with enough diversity, these training sets are typically simulated, an approach called \textit{virtually supervised learning} in, e.g., \cite{foy2021mean,srivastava2022realistic}. For those applications on which the effect of reverberation is detrimental, such as speaker localization or multi-microphone speech enhancement, the use of room acoustics simulators is widespread since it allows physical modeling of the room impulse response (RIR) from the source to the receivers, which includes the inter-channel differences utilized by the methods as well as reverberation. Depending on the type of simulation method used, diverse data can be generated in terms of source and receiver position, acoustic material characteristics, or room geometry, which is necessary for the generalization of the methods to a wide range of real acoustic scenes.

Among the various room acoustics simulators under use, the most widespread by far rely on the image source method (ISM) for shoebox rooms \cite{allen1979image, peterson1986simulating}. Its implementation simplicity and speed allow rapid generation of RIRs for thousands of rooms with randomized dimensions, wall absorption coefficients, and source-receiver positions. More elaborate geometrical \cite{siltanen2007room} or wave-based room acoustics simulators \cite{bilbao2013modeling}, that allow complex geometries and more accurate modeling of propagation effects, are not currently fast enough for this. In the context of supervised learning, they have been used to pre-compute a few large-scale single-channel datasets covering a limited set of conditions only \cite{chen2022soundspaces, tang2022gwa}. Even though shoebox ISM simulation employs stronger acoustical simplifications than more advanced room acoustics simulation methods, it has proven useful in training localization models that then perform adequately well on real datasets \cite{chakrabarty2019multi, adavanne2018sound, perotin2019crnn, diaz2020robust, nguyen2020robust}. Most of these studies are using the simplest form of shoebox ISM, namely, broadband omnidirectional sources and receivers and a global wall- and frequency-independent absorption coefficient, which is also the fastest to simulate. More realistic simulation conditions, such as directional sources and receivers and frequency-dependent surface absorption profiles, can be integrated into shoebox ISM with little computational overhead, at the expense of more complex implementation and specification of simulation parameters. Very few studies perform simulations under these more realistic conditions, except when specialized multichannel receivers are used, e.g., spherical \cite{koyama2022spatial}, Ambisonic \cite{adavanne2018sound, perotin2019crnn}, or binaural \cite{ding2020joint, gaultier2017vast}.

Despite the widespread use of shoebox ISM-based simulators for training speaker localization systems, the effect of integrating more realistic simulation conditions at training time on the localization performance on real recordings at test time has hardly been studied. Receiver directivity does not only concern specialized multichannel receivers, but also common arrays of omnidirectional microphones above some frequency, due to the microphone mounting or the size of the capsule. Source directivity seems also crucial considering that most methods focus on speech signals, with human speakers being highly directive \cite{gonzalez2022near}. Realistic surface absorption profiles can also have a drastic effect on the reverberation reaching the microphones, in terms of spatial distribution and power spectrum. A previous work by the authors
showed a significant performance increase on an acoustic parameter estimation task when measured source and receiver directivities and a natural distribution of frequency-dependent absorption coefficients were integrated in the simulations \cite{srivastava2022realistic}. In the context of speaker localization, we are only aware of one recent study \cite{gelderblom2021synthetic} which analyses the effects of speaker directivity and diffuse late reverberation modeling in the simulations. While diffusion showed no significant impact, the source directivity was found to have a positive impact when testing on speech signals convolved with measured directive RIRs.

In this work, we claim that more realistic simulation at training time can significantly improve real-world localization performance, in a wide variety of scenarios. Results on three multichannel datasets of real human speakers captured in various rooms and with various microphone arrays support this claim. The effects of realistic source and receiver directivities and surface absorption profiles at training time are carefully studied in isolation and in combination for each dataset.
%Additionally, effects of simulation realism on training convergence are discussed.

The structure of the rest of the paper is as follows. Section~\ref{sec:simulation} introduces the \emph{naive} and \emph{advanced} ISM-based simulation modes and Section~\ref{sec:real_set} introduces the real test sets considered in this paper. The experimental setup is described in Section~\ref{sec:training} and the results are analyzed in Section~\ref{sec:expe}. We conclude in Section~\ref{sec:print}.

\section{Image Source Simulation}
\label{sec:simulation}
\subsection{Generalized Image Source Method}
Simulation of multichannel reverberant speech can be expressed as
\begin{equation}  
\label{eq:signal_model}
\mathbf{x}[t] = (\mathbf{h} * s)[t] + \mathbf{n}[t]    
\end{equation}
where $*$ denotes convolution, $\mathbf{h}(t) = [h_1(t),...,h_M(t)]^\mathrm{T}$ the vector of RIRs from the source to $M$ microphones, $\mathbf{n}(t)= [n_1(t),...,n_M(t)]^\mathrm{T}$ additive noise and $s$ the dry source signal. RIR generation based on the ISM for shoebox geometries requires at least the dimensions of the (cuboid) room, the source and receiver coordinates, and a global wall absorption coefficient.
An extended version of the ISM that takes into account frequency-dependent walls, propagation, and directivity effects can be expressed in the frequency domain as
\begin{align}
    \widehat{\mathbf{h}}(f)=\sum_{k=0}^{K}\,&\frac{\exp(-\jmath 2 \pi f r_k/cF_{\textrm{s}})}{r_{k}} \cdot d_{\textrm{air}}(r_k, f)\cdot d_{k}(f)\nonumber\\
    &\cdot \widehat{g}_{\textrm{src}}(-\widetilde{\mathbf{r}}_k,f)\cdot  
    \widehat{\mathbf{g}}_{\textrm{mic}}(\widetilde{\mathbf{r}}_k,f).
    \label{eq:transfer_function}
\end{align}
% \begin{equation}
%     \widehat{\mathbf{h}}(f)=\sum_{k=0}^{K}\,\frac{\exp(\frac{-j 2 \pi f r_k}{cF_{\textrm{s}}})}{r_{k}}  d_{\textrm{air}}(r_k, f) d_{k}(f)
%     \widehat{g}_{\textrm{src}}(-\widetilde{\mathbf{r}}_k,f) 
%     \widehat{\mathbf{g}}_{\textrm{mic}}(\widetilde{\mathbf{r}}_k,f).
% \end{equation}
%Here $\widehat{h}(f)$ is the frequency domain representation of $h$ for frequency $f$ in Hz. 
Here, $c$ is the speed of sound and $F_{s}$ is the sampling frequency. $\mathbf{r}_k$ is the position vector of the $k$-th image source w.r.t.\ the microphone array center, $r_k = ||\mathbf{r}_k||$ is the image-source-to-array distance and $\widetilde{\mathbf{r}}_k=\mathbf{r}_k/r_k$ is the direction unit vector. $d_{\textrm{air}}(r, f)$ is the distance-dependent air attenuation while $d_{k}(f)$ is the compound reflection coefficient of all the surfaces that the $k$-th image has been reflected from. $\widehat{\mathbf{g}}_{\textrm{mic}}(\widetilde{\mathbf{r}})$ denotes the vector of $M$ microphone directivity responses defined with their phase center at the array center, for direction-of-arrival (DOA) $\widetilde{\mathbf{r}}$.  $\widehat{g}_{\textrm{src}}(-\widetilde{\mathbf{r}})$ denotes the source directivity response for direction-of-departure $-\widetilde{\mathbf{r}}$. An alternative implementation of the multichannel receivers can instead utilize individual positions, distances, and DOAs for each microphone in the array, and integrate its local directivity excluding inter-channel array propagation effects. That implementation is suitable for open arrays of directional microphones of known directivity, but it is not suitable for more complex directional arrays that include scattering effects, such as spherical arrays or head-related transfer functions.

\subsection{Advanced v/s Naive Simulation}
In the following, we distinguish the \emph{naive} mode of ISM simulation often used in practice for model training, where \textit{a)} wall absorption is assumed to be frequency-independent and equal for all 6 room surfaces, i.e., $d_k(f) = d^{o_k}$ with $o_k$ being the reflection order and \textit{b)} sources and receivers are assumed to be omnidirectional, i.e., $\widehat{g}_{\textrm{src}}(\widetilde{\mathbf{r}},f) = \widehat{\mathbf{g}}_{\textrm{mic}}(\widetilde{\mathbf{r}},f) = 1$. Additionally we define an \emph{advanced} mode of ISM simulation that incorporates more informed choices on the directivity and absorption components. Regarding absorption, the coefficients in 6 octave bands are drawn from a naturally balanced mix of distributions corresponding to reflective and absorptive wall, ceiling, and floor materials, as described in \cite{foy2021mean, srivastava2022realistic}. These coefficients are then interpolated using half-cosine octave bands in the discrete Fourier domain and turned into minimum-phase responses. Note that both modes of simulation are tuned to yield comparable distributions of reverberation time. Regarding source directivity, the spatially interpolated measured directivities of a head-and-torso-with-mouth simulator (Brüel \& Kjaer HATS 4128-C) and two directive loudspeakers (Genelec 8020 and YAMAHA DXR8) taken from the DIRPAT dataset \cite{brandner2018dirpat} are integrated into the simulation. Regarding microphone array directivities, scenario-based informed choices are made, as detailed in Section \ref{sec:training}. These extensions of the ISM are detailed in \cite{srivastava2022realistic} and are currently available as open source code in the branch \texttt{dev/dirpat} of the pyroomacoustics simulator \cite{scheibler2018pyroomacoustics}. 

%[X] and are currently available as open source code at \texttt{link}\footnote{These two references to the code have been removed to preserve the authors' anonymity.}.

%\cite{srivastava2022realistic} and are currently available as open source code in the branch \texttt{dev/dirpat} of the pyroomacoustics simulator \cite{scheibler2018pyroomacoustics}.

%Source directivity has a strong impact on DOA estimation as shown in  \cite{gelderblom2021synthetic}, therefore we purpose to add more layers of realism on the simulation of RIRs (\textit{Advanced simulated}). We achieve this by including measured directive patterns from the DIRPAT dataset \cite{brandner2018dirpat} specifically we experimented by adding the following source patterns
%HATS 4128 C, Genelec 8020 and YAMAHA DXR8 similarly for microphone we test with directivity of Eigenmike EM32. To simulate realistic acoustic rooms we inspire from work in \cite{foy2021mean} and sample frequency-dependent distribution of the walls according to the material database.
%Detailed implementation can be found in our previous work \cite{}.

\section{Real Test Sets}
\label{sec:real_set}
To examine the impact of increased ISM realism at training time, we evaluate virtually-supervised localization performance on three real datasets of human speakers with spatio-temporal annotations of their activity and position with respect to the microphone array. The selected publicly available datasets are captured in a variety of rooms, and with a different microphone array each. We focus specifically on datasets featuring real human speakers as opposed to ones generated by convolving dry speech signals with measured RIRs. While the latter are commonly used in the speaker localization literature, the former are closer to real world conditions. This study focuses on the most elementary localization task, namely, single-source DOA estimation in $[0,180]^{\circ}$ from a two-second signal recorded using a single microphone pair.
%We took into account 3 publically available real datasets to test the robustness of the network with respect to the added layer of realism. These datasets include recordings of actual human speakers in a range of acoustic conditions, with diverse array configurations and different microphone directive patterns providing more complexity to the real data. We carefully craft many training sets according to the scenario present in real data.

%\vspace{1mm}
%\noindent \textbf{VoiceHome-2 \cite{bertin2019voicehome}.} 
%
\subsection{VoiceHome-2 \cite{bertin2019voicehome}}
This dataset is specifically made for distant speech processing applications in domestic environments. It consists of short commands for smart home devices in French, collected in reverberant conditions and uttered by 12 native French speakers. The data is recorded in 12 different rooms corresponding to 4 houses, with fully annotated geometry, under quiet or noisy conditions. It is captured by a microphone array consisting of 8 MEMS placed near the corner of a cubic baffle. For this study, a two-channel sub-array with aperture 10.4~cm is selected, and 360 two-second speech recordings in quiet conditions are used.

%Every room consists of recordings with 3 different noise conditions with point-like noise source and 1 quiet condition with no possible noise. We took 2-second speech-activated signal from audio signals corresponding to quiet noise conditions. The annotations describing the position of the cubic array and each speaker are provided in the room frame of reference and have been used to calculate ground truth (GT) Doa.

% microphones, wherein each microphone pair is placed on the sidepanel of the cube.
%\vspace{1mm}
%\noindent \textbf{DIRHA \cite{ravanelli2015dirha}.}
%
\subsection{DIRHA \cite{ravanelli2015dirha}}
This multichannel dataset consists of recordings done in the living room and kitchen of a typical apartment. Microphones in different configurations are placed on the walls and ceiling of the 2 rooms. 
%26 microphones in the living room out of which 3 were linear 2 microphone array with an inter-microphone distance of $0.30m$ while other consists of a DICIT array and a circular microphone array placed on the roof, kitchen consist only of a circular microphone array.
6 native English speakers are chosen to speak sentences taken from the Wall Street Journal news text corpus. Annotations of individual microphone positions and speaker positions are provided. For this study, a wall-mounted two-channel microphone array with aperture 30~cm placed in the living room is selected, and 410 two-second speech recordings from the living room are used. 

% A.P with microphone pairs placed on the walls of the room in different configurations.

%As reflections of IS behind the microphone will be considered null in this scenario. 
% A.P The corpus is made of short commands, newspaper articles, and conversational speech uttered by 24 native English speakers. 

%\vspace{1mm}
%\noindent \textbf{STARSS22 \cite{politis2022starss22}.}
%
\subsection{STARSS22 \cite{politis2022starss22}}
This dataset contains recordings of naturally acted scenes of human interaction with spatio-temporal annotations of events belonging to 13 target classes, of which speech is a dominant one. This corpus is part of the development set for Task 3 of the DCASE Challenge 2022. It was captured in facilities at Tampere University and at Sony, with the recording, annotation, and organization of acoustic scenes kept similar on both sites. The Eigenmike spherical microphone array is used to deliver the dataset in two spatial formats, one of which is a tetrahedral sub-array of omnidirectional microphones mounted on a rigid spherical baffle. The corpus is more challenging than the other two in the sense that speakers are free to move and turn naturally during discussions, and that it contains intentional and unintentional sound events other than speech with diffuse and directional ambient noise at substantial levels. We carefully pre-processed the data to extract 2,100 two-second non-overlapping speech excerpts from microphones 6 and 10 out of the tetrahedral sub-array, with an aperture of 6.8~cm.

The three curated test sets add up to a total of 95 minutes DOA-annotated, two-channel, real human speech recordings.

\section{Scenario-Based Simulation and Training}
\label{sec:training}
\subsection{Model}
To select a state-of-the-art learning-based localization method for this study, we examined the extensive literature review in \cite{grumiaux2022survey}. We selected the convolutional neural network architecture proposed by He et al. \cite{he2018deep}, specifically its most recent version in \cite{he2021neural}, due to its multiple distinguishing features. Namely, it works with arbitrary microphone arrays, it is initially designed for DOA estimation, it has been successfully applied to real speech recordings, it is readily extendable to multiple localization, detection and counting (not addressed in this study), and its architecture is available through open source code. The model is trained over different simulated training sets using the ADAM optimizer
%\cite{kingma2014adam}
with a learning rate of $10^{-4}$ and batches of size 16 for a maximum of 110 epochs, with early stopping on validation sets. We used the same input features as in \cite{he2021neural}, namely, concatenated short-time Fourier transforms with 50\% overlap and 42.7~ms windows, except that all the signals considered in this study are down-sampled to 16~kHz, for consistency. 
\subsection{Scenario Based Training}
For all the simulated training sets considered, the default air absorption model of pyroomacoustics is used for $d_{\mathrm{air}}$ 
%\cite{huopaniemi1997modeling}
and image sources are simulated up to order 20. A total of 40k shoebox rooms of sizes uniformly drawn at random in $[3, 10]\times[3, 10]\times[2, 4.5]$~m are simulated, each containing a source and a two-microphone array placed uniformly at random with a minimum source-array and device-wall distance of $30$~cm. The obtained RIRs are convolved with speech excerpts from the Librispeech corpus \cite{panayotov2015librispeech} according to (\ref{eq:signal_model}), yielding 40k two-second two-channel reverberated speech samples, of which 38k are used for training and 2k for validation. We experimented with supervising the model with sets containing 10k to 60k samples. While a strong performance improvement was observed from 10k to 60k, diminishing returns were hit around 40k. Further improvements may nonetheless be achievable using even larger sets.

Uncorrelated white Gaussian noise and diffuse speech-shaped noise convolved with the late part of a random RIR in the same room are added to the reverberated signals. As detailed in \cite{srivastava2022realistic}, noise levels are tuned based on reference scenarios according to the source and receiver considered. This yields consistent bell-shaped signal-to-noise ratio distributions in the range $[15,75]$~dB with a peak at $40$~dB for all of the training sets considered in this study. While a detailed analysis of the specific impact of noise at training time would be of interest, this is out of the scope of this paper. For each of the three real test sets described in Section \ref{sec:real_set}, one \textit{naive} and one \textit{advanced} simulated training set is built based on the two modes of simulation described in Section \ref{sec:simulation} for walls, sources and receivers. For naive sets, ideal omnidirectional receivers are placed at random inside the room using the same apertures as the ones of their corresponding test sets. For VoiceHome-2, the simulated arrays are identical in the naive and advanced sets. This is because MEMS are known to be close to omnidirectional and the directivity of the VoiceHome-2 array is not available. Moreover, early experiments revealed that using mismatched microphone responses at training time was detrimental to the results, a phenomenon also reported in \cite{srivastava2022realistic}. For DIRHA, the advanced simulation places the arrays \textit{on} the room walls, which is equivalent to simulating microphones with a half-sphere directivity. For STARSS22, the advanced simulation uses the measured directivity pattern of the relevant sub-array of the Eigenmike.\footnote{We used the measured directivity made available by Franz Zotter et al. from Graz University: \url{https://phaidra.kug.ac.at/o:69292}.}%, as provided by Franz Zotter and team 

\label{sec:majhead}

% \begin{table}
% \centering
% \begin{tabular}{l|c|c|c|c}
% \hline
% \multicolumn{1}{c|}{Training + validation size} & 10k & 20k & 40k & 60k \\
% \hline
% $\uparrow$ Recall & $71\%$ & $78\%$ & $85\%$ & $86\%$ \\
% $\downarrow$ MAE $({}^{\circ})$ & $11.4$ & $7.8$ & $5.9$ & $5.4$ \\
% \hline
% \end{tabular}
% \caption{Impact of the number of training + validation (2k) samples (number of simulated rooms) in terms of the results on the VoiceHome-2 real test set, using advanced simulation.}
% \label{tab:Impact_of_size_of_the_TRS}
% \end{table}

\section{Experiments and Results}
\label{sec:expe}

%We train for 3 systems corresponding to 3 real dataset on their respective naive and advanced simulated training sets as explained in Section \ref{sec:training} and thus mention them as naive and advanced training. We present all the results on a metric of recall with a tolerance of $10^{\circ}$ combined with the mean angular error and its 95\% confidence interval calculated on all the provided test samples for each test set.
To evaluate the virtually-supervised localization systems, two complementary metrics are used for each test set, namely, the mean angular error (MAE, in degrees) and the ratio of sources localized with an error less than $10^{\circ}$ (Recall, in $\%$), which showed to be an adequate threshold to prune out outliers. Models trained using naive and advanced simulations are compared to the classical learning-free steered response power with phase transform (SRP-PHAT) localization method, as implemented in \cite{scheibler2018pyroomacoustics}.

%To assess the difficulty of the DOA estimation problem on different simulated training sets and to determine whether naive training generalizes to the advanced simulated data. We performed a cross-comparison study on the simulation scenario of voicehome2 test set corresponding results are shown in Table~\ref{tab:simulated_results}. The results shows two major points.
%
%First, Naive training struggles to perform and generalize when tested on an advanced simulated test set, as the model based on advanced training outperforms the latter by a margin of 6\% on recall and 1.7 degrees of improvement on MAE is observed. Second, the advanced training performs similar to naive training on an unseen naive test set. These points provides a holistic view of the improvement brought by advanced training and
\subsection{Simulated Test Sets}
We start by comparing the three methods on two test sets simulated in naive and advanced modes under the VoiceHome-2 scenario. The results are shown in Table~\ref{tab:simulated_results}. First, while all the methods perform well on the naive test set, with nearly perfect recall achieved by both trained models, their performance drastically drops on the advanced test set. This suggests that the presence of realistic wall, source and receiver responses significantly hardens the localization task, even under identical noise and reverberation-time distributions. We have not found direct evidence of this in prior literature and believe this could provide a helpful guideline to improve the evaluation of localization methods on synthetic datasets. Second, as expected, the learning-based methods strongly outperform the learning-free one and the advanced model generalizes significantly better to advanced conditions than the naive one. Moreover, the former seems to perform nearly as well as the latter on naive conditions, despite the mismatch. This strengthens the evidence that speaker localization is inherently more challenging in more realistic conditions.

\begin{table}[h!]
\centering
\caption{Localization results on naive and advanced simulated test sets following the VoiceHome-2 scenario.}
\begin{tabular}{|l|c|c|c|c|}
\cline{2-5}
\multicolumn{1}{c|}{ } & \multicolumn{4}{c|}{\textbf{Simulated Test Sets}} \\ 
\cline{2-5}
\multicolumn{1}{c|}{ } & \multicolumn{2}{c|}{Naive} & \multicolumn{2}{c|}{Advanced} \\ 
\hline
\textbf{Training} & $\uparrow$ Recall & $\downarrow$ MAE & $\uparrow$ Recall & $\downarrow$ MAE \\
\hline
Naive & $96 \%$ & $2.6^{\circ}$ & $74 \%$ & $8.5^{\circ}$ \\
%\hline
Advanced & $95 \%$ & $3.0^{\circ}$& $80 \%$ & $6.7^{\circ}$\\
\hline
SRP-PHAT & $75\%$ &  $11.1^{\circ}$ & $50\%$& $20.8^{\circ}$\\
\hline
\end{tabular}

\label{tab:simulated_results}
\end{table}

\begin{table*}[t!]
    \centering
        \caption{Localization results on three real test sets achieved by the SRP-PHAT baseline and by the supervised model \cite{he2021neural} trained using various simulation modes. Mean angular errors (MAE) are displayed with their $95\%$ confidence interval. Bold numbers indicate the best
system in each column and the systems statistically equivalent to it. Statistical significance was
assessed using McNemar's test for the Recall metric and $95\%$ confidence intervals over
angular error differences for the MAE metric.}
    \begin{tabular}{l|c|c|c|c|c|c}
\hline
\multicolumn{1}{c|}{\textbf{Real Test Sets $\rightarrow$}} & \multicolumn{2}{c|}{\textbf{VoiceHome-2} \cite{bertin2019voicehome}} & \multicolumn{2}{c|}{\textbf{DIRHA} \cite{ravanelli2015dirha}} & \multicolumn{2}{c}{\textbf{STARS22} \cite{politis2022starss22}}  \\
\hline
 \textbf{Methods} & $\uparrow$ Recall & $\downarrow$ MAE $({}^{\circ})$ & $\uparrow$ Recall & $\downarrow$ MAE $({}^{\circ})$ & $\uparrow$ Recall & $\downarrow$ MAE $({}^{\circ})$ \\
\hline
SRP-PHAT & $70 \%$ & $9.9\pm 1.5$ & $61 \%$ & $15.0 \pm 2.3$ & $45 \%$ & $14.9\pm0.6$ \\ 
Naive Training & $78 \%$ & $7.6 \pm 1.2$ & $77 \%$ & $8.4 \pm 1.4$ & $57 \%$ & $12.9 \pm 0.6$ \\
Advanced Training & $\mathbf{85 \%}$ & $\mathbf{5.8 \pm 0.8}$ & $\mathbf{84 \%}$ & $\mathbf{6.3 \pm 1.0}$ & $61 \%$ & $\mathbf{11.4\pm0.5}$ \\
\hline
\textbf{Ablation study} & \multicolumn{5}{l}{} \\
\hline
without wall realism & $83 \%$ &  $6.2\pm0.8$ & $81 \%$ & $7.5 \pm 1.4$ & $59 \%$ & $12.1 \pm 0.6$ \\
without source realism & $82 \%$ & $7.1 \pm 1.1 $ & $80 \%$ & $7.8 \pm 1.2$ & $\mathbf{63 \%}$ & $\mathbf{11.4\pm0.6}$ \\
without receiver realism & N/A & N/A & $78 \%$ & $8.3 \pm 1.5$ & $53 \%$ & $13.4\pm0.6$  \\

% True & $0.25$ & $0.34$ & $123.02$ & $80.14$ \\
%est $\bar{\alpha}$ & & 0.35& 0.33&0.43 & 0.47& \\
\hline

\end{tabular}

    \label{tab:real_results}
\end{table*}

\subsection{Real Test Sets}
The three methods are compared on the three real datasets. As can be seen in the top part of Table~\ref{tab:real_results}, advanced training significantly outperforms naive training by 4 to 7 recall points and $2^{\circ}$ MAE margins across all three datasets, despite using the exact same network architecture. It also largely outperforms the classical SRP-PHAT baseline by 15 to 23 recall points and $3^{\circ}$ to $9^{\circ}$ MAE margins. As predicted in Section~\ref{sec:real_set}, the STARS22 dataset proved most challenging for all the methods. Note that even under the quiet and static conditions of VoiceHome-2 and DIRHA, the baseline results are far from perfect. This shows that two-channel DOA estimation remains a challenging task in real-world settings.

The second half of Table~\ref{tab:real_results} presents an ablation study on the proposed advanced simulation strategy. It reveals that removing any of the three considered layers of realism results in noticeable performance loss. One exception is the use of measured source directivity on STARS22, which does not seem to affect performance. One explanation could be that the human speakers in STARS22 perform significant head rotations, which is not modeled by our framework. The use of a measured array directivity seems to have the strongest impact on this dataset, an observation which we have not found in previous literature for such a simple two-element array. On the other two datasets, the positive impact of source directivity previously reported in \cite{gelderblom2021synthetic} is confirmed, while realistic wall absorptions seem to offer a comparable boost in performance. This is new to the best of our knowledge, and may be explained by the presence of diverse real-world rooms in these datasets. The Python code to reproduce these experiments from training data simulation to evaluation is available here: \texttt{\url{github.com/prerak23/Dir_SrcMic_DOA}}.
%\url{github.com/prerak23/Dir_SrcMic_DOA}.
%\footnote{Link removed to preserve the authors' anonymity.}
%We trained our system on naive and advanced training setups, whereas each one of them is specifically simulated for a particular test set scenario mentioned in section 3.  To verify if the training based on naive simulation generalizes to the data simulated on advanced simulation and to assess the difficulty of DOA estimation task, we performed a cross-comparison study on the simulation scenario of the VoiceHome-2 dataset. The results are shown in table 2 and it clearly states two major points. First, Training based on naive simulation struggles to perform as expected when tested on an advanced simulated test set as the model trained on advanced simulation outperforms by a margin of 6\% on recall and 1.7 degrees of improvement on MAE is observed. Second, model trained on advanced simulations performs the same as the other on an unseen naive test scenario for the model trained on advanced simulation. This points provides and holistic view and similar trends are observed in when we test the naive and advanced trained systems on their respective real test sets. The advanced training , the inclusion of source, microphone and wall realism for VoiceHome-2 and DIRHA while 

\section{Conclusion}
\label{sec:print}
This paper revealed that simulating realistic wall absorption and source/receiver directivities at training time can significantly boost the performance of a virtually supervised speaker localization model across a large test corpus, featuring real human speakers and a variety of microphone arrays and rooms. While these aspects have been mostly ignored in the speaker localization literature, we argue that they can critically benefit both the evaluation of models and their real world applicability. Several research avenues are left to explore. Our preliminary findings revealed that the results are sensitive to the noise distribution at training time, calling for a careful dedicated study of such effects. We also observed that the validation learning curves of models trained on advanced simulation tended to peak earlier than the naive ones. This leads us to believe that there is still room for improvement on the reported results, e.g., by enriching the diversity of directivity profiles through data augmentation. Finally, the inclusion of source and receiver movements at training time or the use of more efficient stochastic late reverberation models constitute worthwhile research directions.

\section{Acknowledgements}
%\textit{Removed to preserve the authors' anonymity}.
This work was made with the support of the French National Research Agency through project HAIKUS “Artifical Intelligence applied to augmented acoustic scenes” (ANR-19-CE23-0023). Experiments presented in this paper were carried out using the Grid’5000 testbed, supported by a scientific interest group hosted by Inria and
including CNRS, RENATER and several Universities as well as
other organizations (see https://www.grid5000.fr).


\bibliographystyle{IEEEtran}
\small
\bibliography{mybib}

\end{document}